\documentclass[twocolumn,showpacs,preprintnumbers,amsmath,amssymb]{revtex4}

\usepackage{graphicx}
\usepackage{dcolumn}
\usepackage{bm}

\begin{document}

\preprint{APS/123-QED}

\title{Field-induced quantum magnetism in the verdazyl-based charge-transfer salt $[$$o$-MePy-V-($p$-Br)$_2]$FeCl$_4$}

\author{Y. Iwasaki$^1$, T. Kida$^2$, M. Hagiwara$^2$, T. Kawakami$^3$, Y. Kono$^4$, S. Kittaka$^4$, T. Sakakibara$^4$, Y. Hosokoshi$^1$, and H. Yamaguchi$^1$}
\affiliation{
$^1$Department of Physical Science, Osaka Prefecture University, Osaka 599-8531, Japan \\ 
$^2$Center for Advanced High Magnetic Field Science (AHMF), Graduate School of Science, Osaka University, Osaka 560-0043, Japan\\
$^3$Department of Chemistry, Osaka University, Toyonaka, Osaka 560-0043, Japan\\
$^4$Institute for Solid State Physics, the University of Tokyo, Chiba 277-8581, Japan\\}

Second institution and/or address\\
This line break forced

\date{\today}

\begin{abstract}
We successfully synthesized a verdazyl-based charge-transfer salt $[$$o$-MePy-V-($p$-Br)$_2]$FeCl$_4$, which has an $S_{\rm{V}}$=1/2 on the radical $o$-MePy-V-($p$-Br)$_2$ and an $S_{\rm{Fe}}$=5/2 on the FeCl$_4$ anion.
$Ab$ $initio$ molecular orbital calculations indicate the formation of an $S_{\rm{V}}$=1/2 honeycomb lattice composed of three types of exchange interaction with two types of inequivalent site. 
Further, the $S_{\rm{V}}$=1/2 at one site is sandwiched by $S_{\rm{Fe}}$=5/2 spins through antiferromagnetic (AF) interactions. 
The magnetic properties indicate that the dominant AF interactions between the $S_{\rm{V}}$ = 1/2 spins form a gapped singlet state, and the remaining $S_{\rm{Fe}}$ = 5/2 spins cause an AF order.
The magnetization curve exhibits a linear increase up to approximately 7 T, and an unconventional 5/6 magnetization plateau appears between 7 T and 40 T.
We discuss the differences between the effective interactions associated with the magnetic properties of the present compound and ($o$-MePy-V)FeCl$_4$.
We explain the low-field linear magnetization curve through a mean-field approximation of an $S_{\rm{Fe}}$ = 5/2 spin model.
At higher field regions, the 5/6 magnetization plateau and subsequent nonlinear increase are reproduced by the $S_{\rm{V}}$ = 1/2 AF dimer, in which a particular internal field is applied to one of the spin sites. 
The ESR resonance signals in the low-temperature and low-field regime are explained by conventional two-sublattice AF resonance modes with easy-axis anisotropy.
These results demonstrate that exchange interactions between $S_{\rm{V}}$ = 1/2 and $S_{\rm{Fe}}$ = 5/2 spins in $[$$o$-MePy-V-($p$-Br)$_2]$FeCl$_4$ realize unconventional magnetic properties with low-field classical behavior and field-induced quantum behavior. 
\end{abstract}

                                           
\maketitle
\section{INTRODUCTION}
In recent decades, composite magnetic materials composed of organic radical and inorganic molecules have been studied extensively because these materials have the potential to realize a unique crystal, magnetic, and electronic structure that cannot be seen in conventional organic or inorganic materials~\cite{Lahti1999,Miller2001,Yamaguchi1990,Masuda2009_1}. 
For instance, in the organic conductor $\lambda$-(BETS)$_2$FeCl$_4$, the magnetic-field-induced superconductivity is realized through the interaction between $\pi$ electron in the organic molecule and the $3d$ electron in the FeCl$_4$ anion~\cite{uji_nature}. 
In $\kappa$-(BETS)$_2$FeBr$_4$ with similar crystal structure, the zero-field ground state is superconducting, and magnetic fields induce another superconducting phase~\cite{Kobayashi_2000, Konoike_2004}. 
The organic salts $\kappa$-(BDH-TTP)$_2$FeX$_4$ (X = Br, Cl) in the low-field and low-temperature regions exhibit a steep negative magnetoresistance caused by a spin-canting transition via the $\pi$-$d$ interaction~\cite{Sugii_2013_1, Sugii_2013_2, Sugii_2014}.
Magnetic studies on the salt (NNDPP)FeBr$_4$ have demonstrated that the interactions between spins on the organic radical NNDPP and FeBr$_4$ anion induce ferrimagnetic behavior~\cite{Masuda2009_2}.

In our previous work, we demonstrated that the verdazyl radical can form a variety of unconventional spin systems, including the ferromagnetic-leg ladder, quantum pentagon, and random honeycomb, which have not been realized in conventional inorganic materials~\cite{3Cl4FV,fine-tune, b26Cl2V, a26Cl2V, random}. 
Furthermore, molecular-based complexes with verdazyl radical and $3d$ transition metals demonstrated that the strong coupling between the metal and verdazyl radical spins results in the formation of a hybrid spin in low-temperature regions~\cite{Zn,Mn}.
Recently, we succeeded in synthesizing verdazyl-based charge-transfer salts by combining cationized verdazyl radicals with anions~\cite{square_TCNQ, PF6,FeCl4}.
In the salt ($o$-MePy-V)FeCl$_4$, metal-radical hybrid spin is formed through the dominant intermolecular interactions between the $S$=1/2 and $S$=5/2 spins, which are located on the verdazyl radical and the FeCl$_4$ anion, respectively~\cite{FeCl4}. 
Its magnetic properties indicated that the hybrid spins that are stabilized in low-temperature regions form an effective $S$ = 2 antiferromagnetic (AF) chain.
Furthermore, because the intermolecular interactions between the radical and anion in ($o$-MePy-V)FeCl$_4$ are much smaller than the intramolecular ones in coordinated complexes~\cite{Mn,Vcomp1,Vcomp2,Vcomp3}, the experimental magnetic fields can modulate the coupled spin state between the verdazyl radical and the FeCl$_4$ anion. 
As a result, the effective $S$=1/2 quantum honeycomb lattice accompanied by the quantum magnetism is realized at high-field regions~\cite{FeCl4} 
These results demonstrate that verdazyl-based salts with magnetic anions can make various forms of field-induced quantum magnetism observable at laboratory level, owing to the moderate energy scale of intermolecular interactions between spins on the radical and magnetic anion.

In this paper, we present a new verdazyl-based charge-transfer salt with a magnetic anion.
We successfully synthesized single crystals of $[$$o$-MePy-V-($p$-Br)$_2]$FeCl$_4$ [$o$-MePy-V-($p$-Br)$_2$ = 3-(2-methylpyridyl)-1,5-bis(4-bromophenyl)-verdazyl]. 
$Ab$ $initio$ molecular orbital calculations indicate the formation of an $S$=1/2 honeycomb lattice composed of three types of exchange interactions with two types of inequivalent sites. 
Further, the $S$=1/2 at one spin site is sandwiched by $S$=5/2 spins through antiferromagnetic (AF) interactions. 
The magnetic properties indicate that the dominant AF interactions between the $S$ = 1/2 spins form a gapped singlet state, and the remaining $S$ = 5/2 spins cause an AF order.
As a result, the magnetization curve exhibits a linear increase up to approximately 7 T, and an unconventional 5/6 magnetization plateau appears between 7 T and 40 T.
The ESR resonance signals in the low-temperature and low-field regime are explained by conventional two-sublattice AF resonance modes with easy-axis anisotropy.

\section{EXPERIMENTAL AND NUMERICAL METHOD}
The synthesis of $[$$o$-MePy-V-($p$-Br)$_2]$FeCl$_4$, whose molecular
structure is shown in Fig. 1(a), was performed using a
procedure similar to that for ($o$-MePy-V)FeCl$_4$~\cite{FeCl4}. 
The recrystallization in acetonitrile yielded dark-red crystals.

Single crystal X-ray diffraction (XRD) experiment was performed by using a Rigaku AFC-8R Mercury CCD RA-Micro7 diffractometer with Japan Thermal Engineering XR-HR10K. 
The single crystal XRD data are refined by using the SHELX software~\cite{Xray}. 
The structural refinement was carried out using anisotropic and isotropic thermal parameters for the nonhydrogen atoms and the hydrogen atoms, respectively. 
All the hydrogen atoms were placed at the calculated ideal positions.

The magnetizations were measured using a commercial SQUID magnetometer (MPMS-XL, Quantum Design) down to 1.8 K.
High-field magnetization measurement in pulsed magnetic fields of up to approximately 52 T was conducted using a non-destructive pulse magnet.
The experimental results were corrected for the diamagnetic contribution of $-3.77{\times}10^{-4}$ emu mol$^{-1}$ calculated by the Pascal method.
The specific heat was measured with a commercial calorimeter (PPMS, Quantum Design) using a thermal relaxation method above 1.9 K and a handmade apparatus by a standard adiabatic heat-pulse method with a $^3$He refrigerator down to about 0.3 K. 
The ESR measurements were performed utilizing a vector network analyzer (ABmm) and a superconducting magnet (Oxford Instruments). 
At approximately 10.9, 19.6, and 27.6 GHz, we used laboratory-built cylindrical high-sensitivity cavities.
All above the experiments were performed using small, randomly oriented  single crystals with typical dimensions of 1.0$\rm{\times}$0.6$\rm{\times}$0.3 mm$^3$.

$Ab$ $initio$ MO calculations were performed using the UB3LYP method with the basis set 6-31G(d,p) in the Gaussian 09 program package. 
The convergence criterion was set at 10$^{-8}$ hartree.
For the estimation of intermolecular magnetic interaction, we applied our evaluation scheme that have been studied previously~\cite{MOcal}.

\section{RESULTS}
\subsection{Crystal structure and magnetic model}
The crystallographic data for the synthesized $[$$o$-MePy-V-($p$-Br)$_2]$FeCl$_4$ are summarized in Table I.
The verdazyl ring (which includes four N atoms), the upper two phenyl rings, and the bottom methylpyridyl ring are labeled ${\rm{R}_{1}}$, ${\rm{R}_{2}}$, ${\rm{R}_{3}}$, and ${\rm{R}_{4}}$, respectively.
The crystals contain two crystallographically independent molecules. 
The results of the MO calculations for each $o$-MePy-V-($p$-Br)$_2$ molecule indicate that approximately 60 ${\%}$ of the total spin density is present on ${\rm{R}_{1}}$. 
Further, while ${\rm{R}_{2}}$ and ${\rm{R}_{3}}$ each account for approximately 19 ${\%}$ and 17 ${\%}$ of the relatively large total spin density, ${\rm{R}_{4}}$ accounts for less than 4 ${\%}$ of the total spin density.
Therefore, the intermolecular interactions are caused by the short contacts related to the ${\rm{R}_{1}}$, ${\rm{R}_{2}}$, and ${\rm{R}_{3}}$ rings.
Note that, because this study focuses on the low-temperature magnetic properties, the crystallographic data obtained at 25 K are used hereafter.

The $o$-MePy-V-($p$-Br)$_2$ and FeCl$_4$ molecules have $S_{\rm{V}}$=1/2 and $S_{\rm{Fe}}$=5/2, respectively.
The $ab$ $initio$ MO calculations were performed in order to evaluate the exchange interaction between the spins, and five types of dominant interactions were found, as shown in Figs. 1(b)-(e).
They are evaluated as $J_{1}/k_{\rm{B}}$ = 25.1 K, $J_{2}/k_{\rm{B}}$ = $-5.1$ K, $J_{3}/k_{\rm{B}}$ = $-3.5$ K, $J_{4}/k_{\rm{B}}$ = $8.4$ K, and $J_{5}/k_{\rm{B}}$ = $5.8$ K, which are defined in the Heisenberg spin Hamiltonian given by $\mathcal {H} = J_{n}{\sum^{}_{<i,j>}}\textbf{{\textit S}}_{i}{\cdot}\textbf{{\textit S}}_{j}$, where $\sum_{<i,j>}$ denotes the sum over the neighboring spin pairs.
The molecular pairs associated with the $J_{1}$, $J_{2}$, and $J_{3}$ are between crystallographically independent $o$-MePy-V-($p$-Br)$_2$ molecules and have C-C short contacts of 3.33, 3.21, and 3.77 $\rm{\AA}$, respectively, as shown in Figs. 1(b), (c), and (d).
The $J_{4}$ and $J_{5}$ describe the couplings between one of the $o$-MePy-V-($p$-Br)$_2$ molecules and FeCl$_4$ molecules, as shown in Figs. 1(e).
The $o$-MePy-V-($p$-Br)$_2$ molecules couple two-dimensionally through the $J_{1}$, $J_{2}$, and $J_{3}$ in the $ab$ plane, as shown in Fig. 1(f), and FeCl$_4$ molecules are located between the two-dimensional (2D) layers, as shown in Fig. 1(g).
Figure 2 shows the 2D honeycomb lattice composed of $J_{1}$, $J_{2}$, and $J_{3}$ with $S_{\rm{V}}$=1/2 in the $ab$ plane, where the $S_{\rm{V}}$ on the site indicated by the gray ball is connected with two $S_{\rm{Fe}}$=5/2 through $J_{4}$ and $J_{5}$. 
Considering the symmetry of the crystal structure, there is another  honeycomb lattice with a slightly different pattern, in which $J_{4}$ and $J_{5}$ are inversely connected to $S_{\rm{V}}$.
The two different lattices stack alternately along the $c$ axis. 
The difference between two lattices does not affect the energy state of the spins, which gives rise to the same ground state.
Therefore, those honeycomb lattices are considered to be topologically equivalent, and we regard them as the same system hereafter.

\begin{table}
\caption{Crystallographic data for $[$$o$-MePy-V-($p$-Br)$_2]$FeCl$_4$.}
\label{t1}
\begin{center}
\begin{tabular}{ccc}
\hline
\hline 
Formula & \multicolumn{2}{c}{C$_{20}$H$_{17}$Br$_{2}$Cl$_{4}$FeN$_{5}$}\\
Crystal system & \multicolumn{2}{c}{Monoclinic}\\
Space group & \multicolumn{2}{c}{$P$2$_1$/$a$}\\
Temperature (K) & RT & 25(2)\\
Wavelength ($\rm{\AA}$) & \multicolumn{2}{c}{0.7107} \\
$a (\rm{\AA}$) &  14.830(5) & 14.36(7) \\
$b (\rm{\AA}$) &  17.848(6) & 17.95(8) \\
$c (\rm{\AA}$) &  20.218(8) & 19.70(9)\\
$\beta$ (degrees) &  105.391(7)  & 103.71(6)\\
$V$ ($\rm{\AA}^3$) & 5160(3) &  4933(40) \\
$Z$ & \multicolumn{2}{c}{4} \\
$D_{\rm{calc}}$ (g cm$^{-3}$) & 1.763 & 1.844\\
Total reflections & 8508 & 8142\\
Reflection used & 4245 & 6459\\
Parameters refined & \multicolumn{2}{c}{579}\\
$R$ [$I>2\sigma(I)$] & 0.0812 & 0.0501\\
$R_w$ [$I>2\sigma(I)$] & 0.2354 & 0.1031\\
Goodness of fit & 1.083 & 0.972\\
CCDC &  1865093 & 1865094 \\
\hline
\hline
\end{tabular}
\end{center}
\end{table}

\begin{figure*}[t]
\begin{center}
\includegraphics[width=36pc]{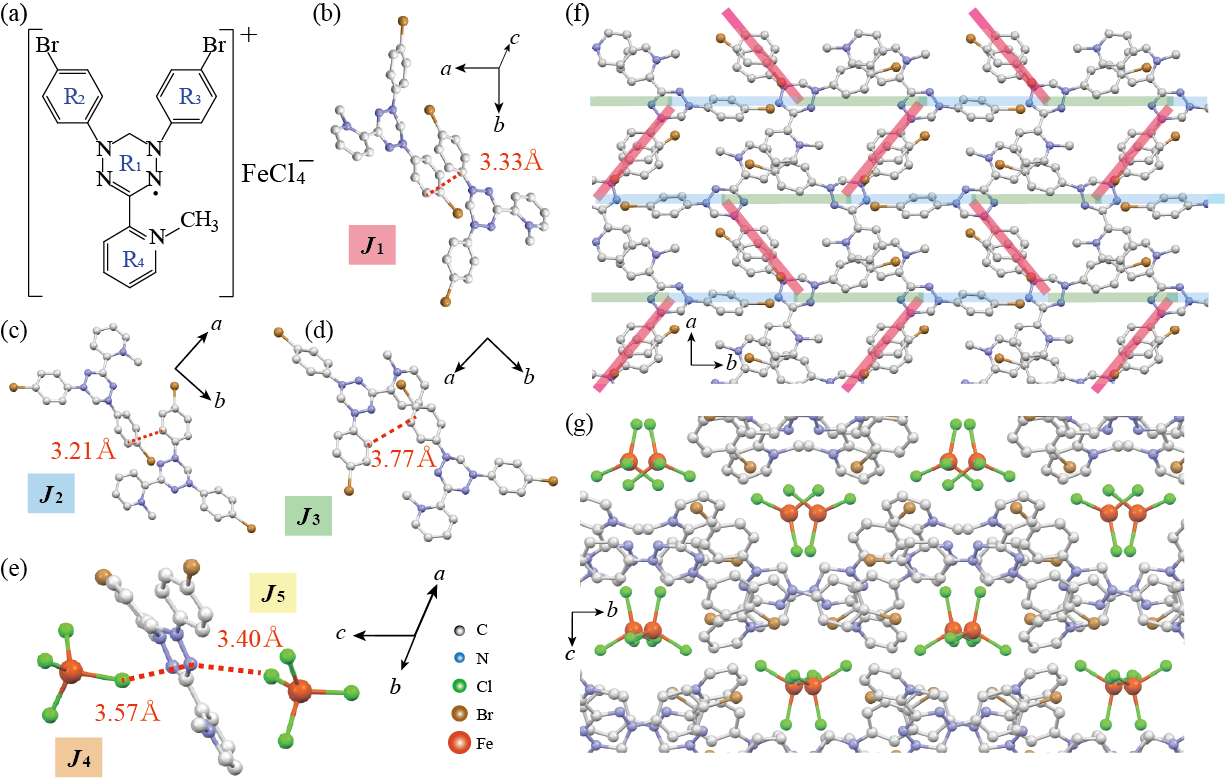}
\caption{(color online) (a) Molecular structure of $[$$o$-MePy-V-($p$-Br)$_2]$FeCl$_4$. Molecular pairs associated with exchange interactios (b) $J_1$, (c) $J_2$, (d) $J_3$, and (e) $J_4$ and $J_5$. Hydrogen atoms are omitted for clarity. The broken lines indicate C-C and N-Cl short contacts. Crystal structure in the (f) $ab$ and (g) $bc$ planes. The lines represent $J_1$, $J_2$ and $J_3$ forming the honeycom lattice.}\label{f1}
\end{center}
\end{figure*}

\begin{figure}[t]
\begin{center}
\includegraphics[width=16pc]{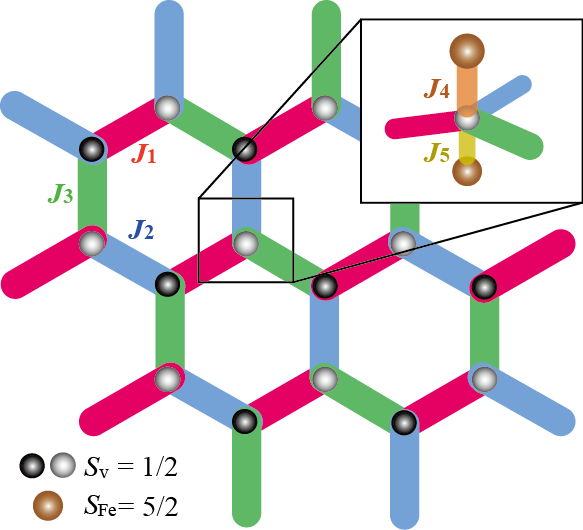}
\caption{(color online) 2D spin model composed of $J_i$ ($i$=1-5) with $S_{\rm{V}}$=1/2 and $S_{\rm{Fe}}$=5/2 in the $ab$ plane. The $J_{1}$, $J_{2}$, and $J_{3}$ form the honeycomb lattice of $S_{\rm{V}}$, and  $S_{\rm{V}}$ on the site shown by the gray ball is sandwiched by $S_{\rm{Fe}}$=5/2 spins through the $J_{4}$ and $J_{5}$, yielding the particular internal field. }\label{f2}
\end{center}
\end{figure}

\subsection{Magnetic susceptibility}　
Figure 3 shows the temperature dependence of the magnetic susceptibility ($\chi=M/H$) at 0.1 and 1.0 T. 
We observe an anomalous change in the temperature dependence at 3.4 K, below which a significant difference between 0.1 and 1.0 T appears.
This behavior indicates that an AF phase transition to a three-dimensional (3D) long-range order (LRO) occurs at $T_{\rm{N}}$ = 3.4 K.
The $\chi T$ decreases with decreasing temperature, indicating dominant contributions of AF interactions, as shown in the inset of Fig. 3.
In the high-temperature region, the value of $\chi T$ approaches $\sim$ 4.7 emu K/mol, which is close to the expected value for the noninteracting $S_{\rm{V}}$=1/2 and $S_{\rm{Fe}}$=5/2 spins.  
The temperature dependence of $\chi T$ is dramatically different from that of ($o$-MePy-V)FeCl$_4$, in which $\chi T$ exhibits a constant of 3.0 emu K/mol for an $S$ = 2 hybrid spin through the strong coupling between $S_{\rm{V}}$=1/2 and $S_{\rm{Fe}}$=5/2 spins~\cite{FeCl4}.
Accordingly, the experimental result of $\chi T$ in the present compound indicates that the exchange interactions between $S_{\rm{V}}$=1/2 and $S_{\rm{Fe}}$=5/2 spins (i.e., $J_{4}$ and $J_{5}$) are relatively weak compared to the dominant interaction between $S_{\rm{V}}$=1/2 spins.

\begin{figure}[t]
\begin{center}
\includegraphics[width=18pc]{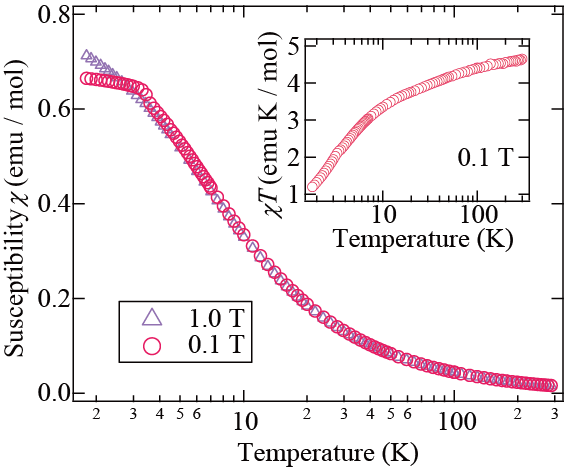}
\caption{(color online) Temperature dependence of magnetic susceptibility ($\chi=M/H$) of $[$$o$-MePy-V-($p$-Br)$_2]$FeCl$_4$ at 0.1 and 1.0 T. The inset shows the temperature dependence of $\chi T$.}\label{f3}
\end{center}
\end{figure}

\subsection{Specific heat}
The experimental results for the specific heat $C_{\rm{p}}$ at zero-field clearly exhibit a $\lambda$-type sharp peak at $T_N$, which is associated with the AF phase transition to the LRO, as shown in Fig. 4.
Although the lattice contribution is not subtracted from the experimental results for specific heat, the magnetic contribution is expected to be dominant in low-temperature regions below $T_N$, as seen in other verdazyl-based materials~\cite{3Cl4FV, b26Cl2V, Zn, 2Cl6FV}.
The entropy $S_{\rm{p}}$ obtained through integration of $C_{\rm{p}}$/$T$ shows that the change associated with the phase transition is close to the total magnetic entropy of $S_{\rm{Fe}}$ = 5/2 ($R$ln6 $\simeq$ 14.9), as shown in the lower inset of Fig. 4.
Therefore, the observed phase transition should originate from the LRO of an effective spin model composed of the $S_{\rm{Fe}}$=5/2.
Because the magnetic entropy of $S_{\rm{V}}$=1/2 is not associated with the phase transition, it is deduced that the strongest AF interaction $J_1$ forms an $S_{\rm{V}}$=1/2 AF dimer with a nonmagnetic singlet state in higher temperature regions.  

As shown in the upper inset of Fig.4, in the low-temperature region below $\sim $0.8 K, $C_{\rm{p}}/T$ shows clear $T$-linear behavior, which suggests the existence of a linear dispersive mode in a 2D AF system.
Thus, we expect that the effective $S_{\rm{Fe}}$ = 5/2 model associated with the phase transition has a quasi-2D character. 
A higher-temperature small shoulder observed at approximately 1.0 K is considered to originate from contributions of some higher-energy dispersive modes.
In magnetic fields, the phase transition temperature decreases with increasing fields, as shown in Fig. 5(a), and the obtained magnetic field dependence of $T_{\rm{N}}$ is shown in Fig. 5(b).
The disappearance of the phase transition at approximately 7 T is considered to correspond to a fully polarized state of the effective $S_{\rm{Fe}}$ = 5/2 model in the low-temperature region, which is consistent with an appearance of a 5/6 magnetization plateau in the following magnetization curve.

\begin{figure}[t]
\begin{center}
\includegraphics[width=18pc]{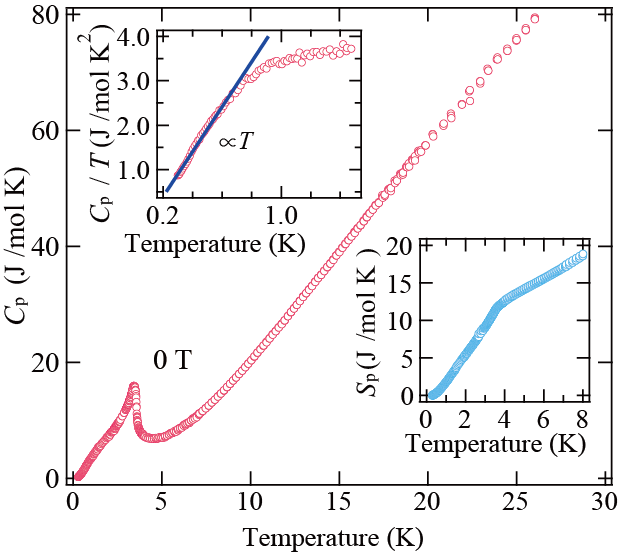}
\caption{(color online) Specific heat $C_{\rm{p}}$ of $[$$o$-MePy-V-($p$-Br)$_2]$FeCl$_4$ at zero-field. The upper and lower insets show the low-temperature part of $C_{\rm{p}}/T$ and the evaluated entropy, respectively. The solid line shows the $T$-linear fit below approximately 0.8 K. }\label{f4}
\end{center}
\end{figure}

\begin{figure}[t]
\begin{center}
\includegraphics[width=18pc]{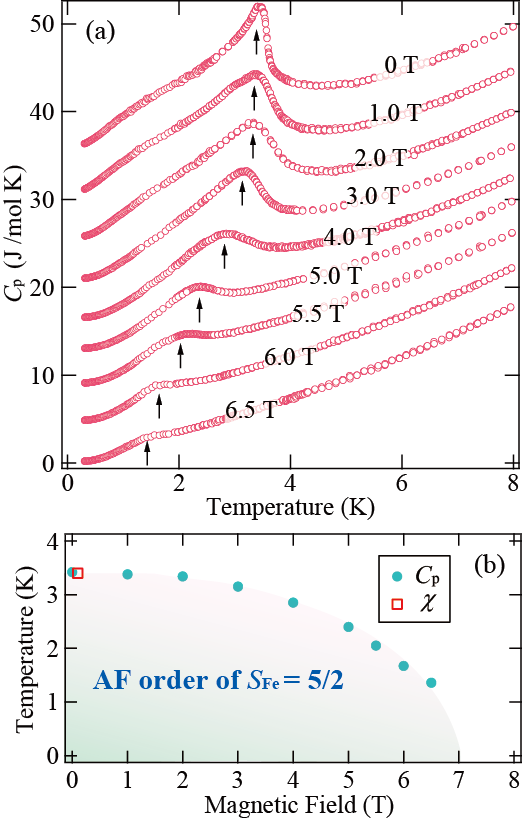}
\caption{(color online) (a) Specific heat $C_{\rm{p}}$ of $[$$o$-MePy-V-($p$-Br)$_2]$FeCl$_4$ in the low-temperature
regions at various magnetic fields. The arrow indicate
the phase transition temperatures. 
For clarity, the values for 0, 1.0, 2.0, 3.0, 4.0, 5.0, 5.5, and 6.0 T have been shifted up by 36, 31, 26, 21, 16, 13, 9.0 and 4.7 J mol$^{-1}$ K, respectively. (b) Magnetic field vs temperature phase diagram showing AF order of $S_{\rm{Fe}}$=5/2 spins. The square and circles indicate the phase
boundaries determined from the magnetic susceptibility and the specific
heat, respectively}\label{f4}
\end{center}
\end{figure}

\subsection{Magnetization curve}
Figure 6 shows the magnetization curve at 1.5 K, which exhibits an almost linear increase with increasing fields up to approximately 7 T.
In terms of the field derivative of the magnetization, we observe a distinct peak at approximately $H_{\rm{SF}}\simeq$ 0.54 T, as shown in the inset of Fig. 6. 
This indicates a spin-flop transition that is caused by a small magnetic anisotropy, which we discuss in the following ESR section. 
The magnetization assumes a 5/6-plateau for fields between 7 and 40 T and then increases again towards saturation at approximately 50 T.
The linear increase observed in the low-field region indicates that the magnetic behavior can be described by a classical system with a large spin size.
Furthermore, the magnetic moment of $5{\mu _B}$/f.u. at the 5/6-plateau phase corresponds to the fully polarized $S_{\rm{Fe}}$=5/2 spins along the field direction.
These characteristics are consistent with the formation of the effective $S_{\rm{Fe}}$ = 5/2 model in the low-temperature and low-field regions. 
In the case of higher-field regions, the observed 5/6-plateau indicates the coexistence of the fully polarized $S_{\rm{Fe}}$=5/2 spins and a singlet state separated from the excited states by an energy gap.
The increase of the magnetization curve toward the saturation exhibits a nonlinear behavior, which reflects the strengths of the quantum fluctuations.
In general quantum spin systems, quantum fluctuations are suppressed by the application of magnetic fields, yielding a nonlinear increase of magnetization curve~\cite{2Cl6FV, 2Cl36F2V, a235Cl3V}.

\subsection{Electron spin resonance}
We performed the ESR measurements in the low-field and low-tempereture regime to examine the ground state of the $S_{\rm{Fe}}$ = 5/2 AF spin lattice.
The frequency dependence of the ESR absorption spectra in the ordered phase is presented in Figs. 7(a) and (b).
As shown in Fig. 7(a), the resonance signals at high frequencies are almost proportional to the external field.
Conversely, those at low frequencies in Fig. 7(b) exhibit broad signals with a number of resonance fields and obviously deviate from the linear field behavior. 
All the resonance fields are plotted in the frequency$-$field diagram, presented in Fig. 8.
Since a zero-field gap of $\sim $15 GHz, which corresponds to the energy scale of $H_{\rm{SF}}$, is expected from the extrapolation of the resonance modes, the observed resonance fields suggest conventional AF resonance modes in an anisotropic two-sublattice model~\cite{FeCl4,a235Cl3V, kittel, MnF2, MnCl3(bipy)_hagiwara, MnCl3bpy}.
The anisotropic energy derived from the dipole$-$dipole interactions is confirmed to induce observable magnetic anisotropy even in isotropic radical sysytems~\cite{FeCl4, a235Cl3V}.

\section{DISCUSSION}
\subsection{Magnetization curve}
Considering the MO calculation and the magnetic properties, the strongest AF interaction $J_1$ is expected to form an $S_{\rm{V}}$=1/2 AF dimer. 
Additionally, at higher field regions, an effective internal field on one of the $S_{\rm{V}}$ sites arises from the fully polarized $S_{\rm{Fe}}$=5/2 spins through the AF $J_{4}$ and $J_{5}$.
Thus, we calculated the magnetization curve for the $S_{\rm{V}}$=1/2 AF dimer coupled through $J_1$ with an effective internal field given by $H_{\rm{in}} = \frac{5}{2}(J_{4}+J_{5})/g{\mu _B}$, which is unusually oriented against the direction of the external field.
The spin Hamiltonian is expressed as
\begin{equation}
\mathcal {H} =J_{1}\textbf{{\textit S}}_{1}{\cdot}\textbf{{\textit S}}_{2}-g{\mu _B}S_{1}^{z}H-g{\mu _B}S_{2}^{z}(H-H_{\rm{in}}),
\end{equation}
where $\textbf{{\textit S}}$ is an $S$ = 1/2 spin operator, ${\mu}$$_B$ is the Bohr magneton, and $H$ is the external magnetic field, and $z$ axis is parallel to the external field direction.
The MO calculation showed that the exchange interactions have the relation $J_{4}/J_{1}$ = 0.33 and $J_{5}/J_{1}$ = 0.23.
Assuming these ratios, we demonstrate the drastic change of magnetization between approximately 40 T and 45 T by using parameters $J_{1}/k_{\rm{B}}$ = 28.3 K, $J_{4}/k_{\rm{B}}$ = 9.3 K, and $J_{5}/k_{\rm{B}}$ = 6.5 K, as shown in Fig. 6.
The obtained parameters are moderately consistent with those evaluated from the MO calculation.
In the actual spin model, finite couplings between the dimers are expected to cause an AF 3D LRO when the energy gap closes by applying magnetic fields.
In the ordered phase, the magnetization curve becomes more gradual compared to that of the isolated dimer owing to interdimer interactions.
The difference between the experimental and calculated results of the magnetization curve should arise from such interdimer contributions. 
Hence, we qualitatively confirm that, above approximately 7 T, the effective spin model in the high-field region can be considered as the $S_{\rm{V}}$=1/2 AF dimer with the particular internal field caused by the fully polarized $S_{\rm{Fe}}$=5/2 spins.  

The $S_{\rm{V}}$=1/2 AF dimer coupled with $J_{1}/k_{\rm{B}}$ = 28.3 K forms a nonmagnetic singlet state at sufficiently high-temperature regions above $T_{\rm{N}}$, and thus the interactions between $S_{\rm{V}}$=1/2 and $S_{\rm{Fe}}$ = 5/2 spins can be omitted to simplify the spin model.
Accordingly, in the low-temperature and low-field regions, the magnetic properties originate from the effective spin model composed of $S_{\rm{Fe}}$ = 5/2, which exhibits phase transition to the AF LRO at $T_{\rm{N}}$ and quasi-2D character in the specific heat.
Since the FeCl$_4$ molecules with $S_{\rm{Fe}}$ are stuck between two radical layers forming the nonmagnetic state (see Fig. 1(g)), the exchange paths between $S_{\rm{Fe}}$ = 5/2 spins in the FeCl$_4$ layer are essential for considering the spin lattice.
The symmetry of the crystal structure indicates that there are two types of FeCl$_4$ layers with similar molecular arrangements.
From the MO calculations, the absolute values of the exchange interactions in both FeCl$_4$ layers are evaluated to be less than 0.5 K.
Considering a strong dependence on the calculation method, those small interactions do not have enough reliability to assume a spin model~\cite{MOseido}. 
Thus, we directly examined the distances between the FeCl$_4$ molecules and found two types of Cl-Cl short contacts less than 5.0 $\rm{\AA}$ in each FeCl$_4$ layer, which form a honeycomb lattice, as shown in Fig. 9.
In consideration of the small energy scale, we assume the other exchange paths between distant sites in order to consider the magnetic properties appropriately in the low-temperature region. 
We then calculated the magnetization curve based on the $S_{\rm{Fe}}$ = 5/2 AF spin lattice using a mean-field approximation assuming the spin Hamiltonian expressed as
$\mathcal {H} = J{\sum^{}_{<ij>}}\textbf{{\textit S}}_i{\cdot}\textbf{{\textit 
S}}_{j}-g{\mu _B}H{\sum^{}_{i}}S_z$,
where $\textbf{{\textit S}}$ is the $S$ = 5/2 spin operator.
The magnetization curve at $T$ = 0 is given by 
$M_{\rm{mean}}=g^{2}{\mu _B}H/2ZJ$,
where $Z$ is the number of nearest-neighbor spins.
Considering the saturation field of approximately 7 T evaluated from the extrapolation of phase boundary, we determined $ZJ/k_{\rm{B}}$ = $1.86$ K, which corresponds to $J/k_{\rm{B}}$ = $0.62$ K assuming the honeycomb lattice ($Z$ = 3).
We obtained good agreement between the experimental and calculated results in the low-field region (as shown in Fig. 6), while there was a slight difference attributed to the finite temperature effect in the experimental results.

\begin{figure}[t]
\begin{center}
\includegraphics[width=20pc]{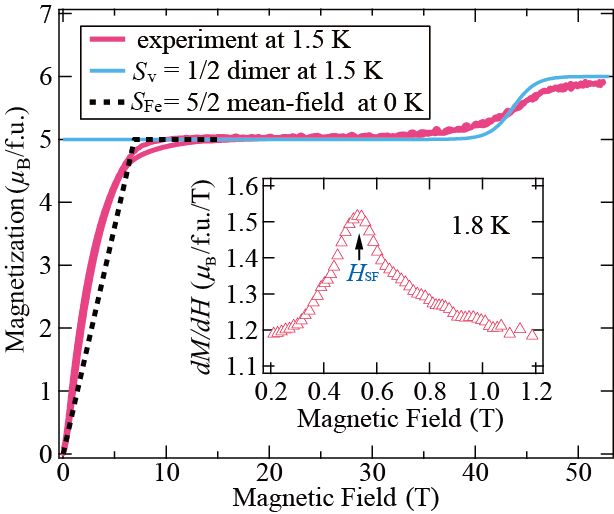}
\caption{(color online) Magnetization curve ${M}$ of $[$$o$-MePy-V-($p$-Br)$_2]$FeCl$_4$ at 1.5 K. The solid and broken lines represent the calculated result for the $S_{\rm{V}}$ = 1/2 AF dimer at the experimental temperature and for the $S_{\rm{Fe}}$ = 5/2 spin model using the mean-field approximation at zero-temperature, respectively. 
The inset shows $dM/dH$ at 1.8 K, and the arrow indicates a sharp peak associated with the spin-flop transition at $H_{\rm{SF}}$.
}\label{f4}
\end{center}
\end{figure}

\begin{figure}[t]
\begin{center}
\includegraphics[width=18pc]{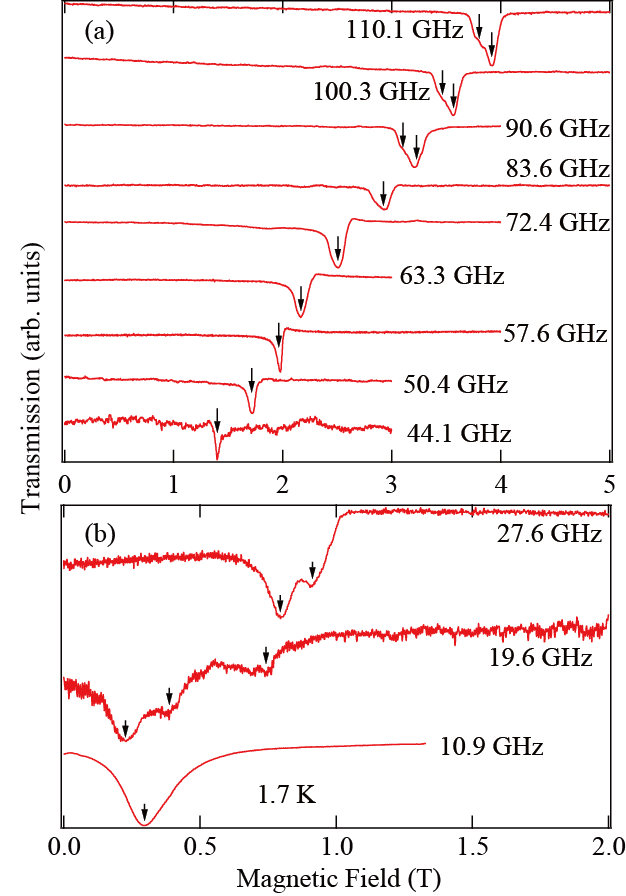}
\caption{(color online) Frequency dependence of ESR absorption spectra of $[$$o$-MePy-V-($p$-Br)$_2]$FeCl$_4$ at 1.7 K for (a) directly detected high-frequencies and (b) low-frequencies measured by cylindrical high-sensitivity cavities. The arrows indicate resonance signals. }
\end{center}
\end{figure}

\begin{figure}[t]
\begin{center}
\includegraphics[width=18pc]{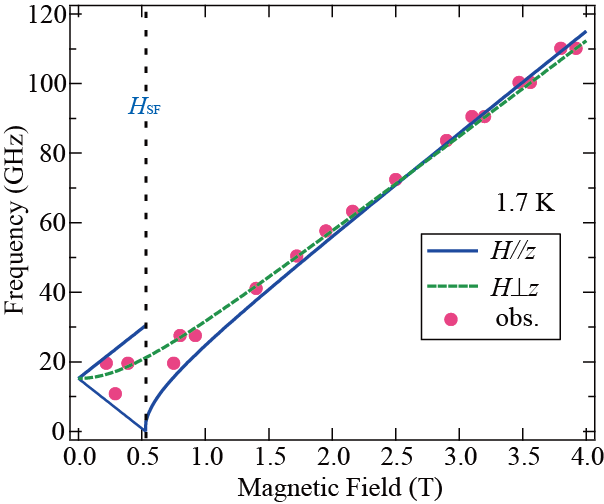}
\caption{(color online) Frequency-field plot of the ESR resonance fields at 1.7 K. The solid lines indicate the calculated AF resonance modes for $H{\parallel}z$ and $H{\perp}z$. The discontinuous changes of the lines correspond to the spin-flop transition at $H_{SF}$.
}
\end{center}
\end{figure}

\begin{figure}[t]
\begin{center}
\includegraphics[width=20pc]{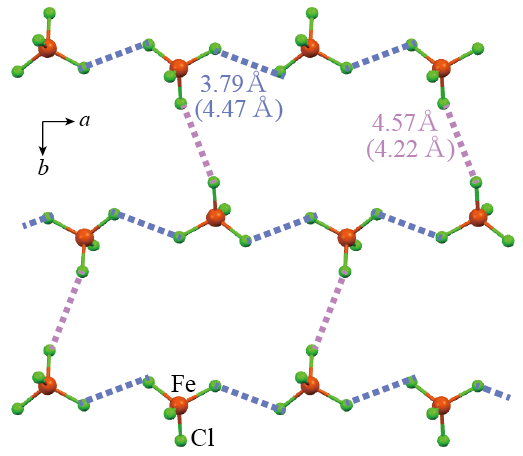}
\caption{(color online) One of the FeCl$_4$ layers of $[$$o$-MePy-V-($p$-Br)$_2]$FeCl$_4$ in the $ab$ plane. The broken lines indicate Cl-Cl short contacts. The corresponding distances for the other FeCl$_4$ layer are in parentheses. 
}
\end{center}
\end{figure}

\subsection{ESR resonance modes}
We analyzed the observed ESR modes in terms of a mean-field approximation assuming the $S_{\rm{Fe}}$ = 5/2 AF lattice with an easy-axis anisotropy. 
Thus, the spin Hamiltonian is expressed as
\begin{equation}
\mathcal {H} =J{\sum^{}_{<i,j>}}\textbf{{\textit S}}_{i}{\cdot}\textbf{{\textit S}}_{j}-D{\sum^{}_{i}}(\mbox{$S$}^{z}_{i})^2-g{\mu _B}{\sum^{}_{i}}\textbf{{\textit S}}_i{\cdot}\textbf{{\textit H}},
\end{equation}
where $D$ is on-site anisotropy ($D>0$), and $\textbf{{\textit S}}$ is an $S$ = 5/2 spin operator.
As the spin structure is described by the two-sublattice model, the free energy $F$ is expressed in the following form, using the mean-field approximation:
\begin{equation}
F=A{\textbf{{\textit M}}_{1}}{\cdot}{\textbf{{\textit M}}_{2}}-K\{({\mbox{$M$}^{z}_{1}})^{2}+({\mbox{$M$}^{z}_{2}})^{2}\}-({\textbf{{\textit M}}_{1}}+{\textbf{{\textit M}}_{2}}){\cdot}{\textbf{{\textit H}}}, 
\end{equation}
where $A$ and $K$ are given by
\begin{equation}
A=\frac{2}{N}\frac{ZJ}{(g{\mu _B})^2},\>\>
K=\frac{2}{N}\frac{D}{(g{\mu _B})^2},
\end{equation}
and ${\textbf{{\textit M}}_1}$ and ${\textbf{{\textit M}}_2}$ are the sublattice moments expressed as  
\begin{equation}
{\textbf{{\textit M}}_i} = \frac{N}{2}g{\mu _B}{\textbf{{\textit S}}_i}. 
\end{equation}
Here, $N$ is the number of spins, and ${\textbf{{\textit S}}_i}$ is the spin on the $i$-th sublattice ($i$=1,2). 
We derive the resonance conditions by solving the equation of motion
\begin{equation} 
{\partial}{\textbf{{\textit M}}_i}/{\partial}t=\gamma[{\textbf{{\textit M}}_i}\times{\textbf{{\textit H}}_i}],
\end{equation}
where $\gamma$ is the gyromagnetic ratio and ${\textbf{{\textit H}}_i}$ is the mean field applied on the $i$-th sublattice moment given by
\begin{equation} 
{\textbf{{\textit H}}_i}=-{\partial}F/{\partial}{\textbf{{\textit M}}_i}.
\end{equation}
To solve the equation of motion, we use a method for the analysis of ABX$_3$-type antiferromagnets~\cite{tanaka}.
Assuming precession motion of the sublattice moments around those equilibrium directions, we utilize the following expressions, which represent the motion of the $i$-th sublattice moment:
\begin{equation}
{\textbf{{\textit M}}_i}=({\Delta}M_{i\Acute{x}}\exp(i{\omega}t),{\Delta}M_{i\Acute{y}}\exp(i{\omega}t),|{\textbf{{\textit M}}_i}|),
\end{equation}
where ${\Delta}M_{i\Acute{x}},{\Delta}M_{i\Acute{y}}{\ll}|{\textbf{{\textit M}}_i}|$, and $\Acute{x}$, $\Acute{y}$ and $\Acute{z}$ are the principal axes of the coordinate system on each sublattice moment.
The $\Acute{z}$-axis is defined as being parallel to the direction of each sublattice moment, and the $\Acute{x}$- and $\Acute{y}$-axes are perpendicular to the $\Acute{z}$-axis.

The spins are aligned along the easy-axis ($z$ axis) under zero-field conditions, and the discontinuous spin-flop phase transition occurs at $H_{\rm{SF}}$ for $H{\parallel}z$.
The value of $H_{\rm{SF}}$ is expressed as
\begin{equation} 
H_{\rm{SF}}=\frac{5\sqrt{ZJD-D^{2}}}{g{\mu _B}},
\end{equation}
which corresponds to the zero-field energy gap of resonance modes.
Above $H_{\rm{SF}}$, the two sublattices are tilted with respect to the field direction with equivalent angles, while for the other principal axes, where the external fields are applied perpendicular to the easy-axis, the two sublattices are tilted from the easy-axis with equivalent angles along each field direction.
The angles between the sublattice moment and the external field for both directions can then be determined by minimizing the free energy. 
Then, the ${\omega}$ values are obtained by solving eq.(5) numerically.
The calculated results obtained here demonstrate typical AF resonance modes with an easy-axis anisotropy in a two-sublattice model.
Since our experiments were performed using small, randomly oriented  single crystals, the resonance fields for all of the principal axes are expected to have been detected in our experiments.
By using $ZJ_/k_{\rm{B}}$ = 1.86 K evaluated from the analysis of the magnetization curve, we obtained a good fit between the experimental and calculated values with $D/k_{\rm{B}}$ = 0.012 K, $g_{\parallel}$ = 2.05(3) for $H{\parallel}z$, and $g_{\perp}$ = 2.00(2) for $H{\perp}z$, as shown in Fig. 8. 

\section{Summary}
We have succeeded in synthesizing single crystals of the verdazyl-based charge-transfer salt $[$$o$-MePy-V-($p$-Br)$_2]$FeCl$_4$. 
$Ab$ $initio$ MO calculations indicated the formation of an $S_{\rm{V}}$=1/2 honeycomb lattice composed of three types of exchange interaction with two types of inequivalent site. 
At one spin site, the $S_{\rm{V}}$=1/2 is sandwiched by $S_{\rm{Fe}}$=5/2 spins through AF interactions. 
The magnetic susceptibility and specific heat indicated the phase transition to the AF order, and the low-temperature magnetization curve exhibited an unconventional 5/6 magnetization plateau.
These observed behaviors indicated that the dominant AF interactions between the $S_{\rm{V}}$ = 1/2 spins form a gapped singlet state, and the remaining $S_{\rm{Fe}}$ = 5/2 spins cause the AF order.
We described the linear magnetization curve below 7 T using the mean-field approximation of an $S_{\rm{Fe}}$ = 5/2 spin model.
For the magnetization curve at higher field regions, the 5/6 magnetization plateau and subsequent nonlinear increase were demonstrated by the $S_{\rm{V}}$ = 1/2 AF dimer. 
The ESR resonance signals in the low-temperature and low-field regime suggested conventional two-sublattice AF resonance modes with an easy-axis anisotropy.  
We explained the obtained ERS resonance signals assuming the effective $S_{\rm{Fe}}$ = 5/2 spin model by using the mean-field approximation and evaluated magnetic parameters.
These results thus demonstrate that exchange interactions between $S_{\rm{V}}$ = 1/2 and $S_{\rm{Fe}}$ = 5/2 in $[$$o$-MePy-V-($p$-Br)$_2]$FeCl$_4$ realize unconventional magnetic properties with low-field classical behavior and field-induced quantum behavior. 
Verdazyl-based charge-transfer salts with magnetic anions provide a means to observe various types of field-induced quantum magnetism in experimentally accessible magnetic fields.

\begin{acknowledgments}
This research was partly supported by Grant for Basic Science Research Projects from KAKENHI (No. 15H03695, No. 15K05171, and No. 17H04850) and the Matsuda Foundation.
A part of this work was carried out at the Center for Advanced High Magnetic Field Science in Osaka University under the Visiting Researcher's Program of the Institute for Solid State Physics, the University of Tokyo, and the Institute for Molecular Science.
\end{acknowledgments}


\end{document}